\documentclass[aps,showpacs,preprintnumbers]{revtex4}
\usepackage{amssymb}
\usepackage{amsmath}
\usepackage{bm}

\input{tcilatex}

\begin{document}

\title{Antiparticle Contribution in the Cross Ladder Diagram for Two Boson
Propagation in the Light-front.}
\author{J.H.O. Sales}
\affiliation{Instituto de Ci\^{e}ncias Exatas, Universidade Federal de Itajub\'{a}, CEP
37500-000, Itajub\'{a}, MG, Brazil}
\author{A.T. Suzuki}
\affiliation{Instituto de F\'{\i}sica Te\'{o}rica/UNESP, Rua Pamplona, 145 CEP 01405-900 S%
\~{a}o Paulo, SP, Brazil}
\date{\today }

\begin{abstract}
In the light-front milieu, there is an implicit assumption that the vacuum
is trivial. By this \ " triviality " is meant that the Fock space of
solutions for equations of motion is sectorized in two, one of positive
energy $k^{-}$ and the other of negative one corresponding respectively to
positive and negative momentum $k^{+}$. It is assumed that only one of the
Fock space sector is enough to give a complete description of the solutions,
but in this work we consider an example where we demonstrate that \textsl{%
both} sectors are necessary.

.
\end{abstract}

\maketitle

\address{ 27607.}









\section{Introduction}

In this work we explore the concept of covariant quantum propagator written
down in terms of light front coordinates and obtain the propagator and
Green's function in the light-front for a time interval $x^{+}$ where $%
x^{+}=t+z$, is the light-front \textquotedblleft time\textquotedblright . In
principle, this is equivalent to the canonical quantization in the light
front \cite{20,21,22,33,34,35,36}. Kogut and Soper \cite{22} also makes use
of this way of constructing quantities in the light front: starting with $4$%
-dimensional amplitudes or equations, they integrate over $k^{-}=k^{0}-k^{3}$%
, which plays the role of \textquotedblleft energy\textquotedblright\ and
corresponds to processes described by amplitudes or equations in
\textquotedblleft time\textquotedblright\ $x^{+}$. With this, the relative
time between particles disappear and only the global propagation of
intermediate state is allowed. The global propagation of the intermediate
state is the \textquotedblleft time\textquotedblright\ translation of the
physical system between two instants $x^{+}$ and $x^{^{\prime }+}$.

We specifically consider the case of two scalar bosons propagating forward
in time. It has already been shown that for this case we can have exchange
of two intermediate bosons via ladder diagrams or crossed ladder diagrams.
For the ladder diagram we do not observe any new features, but for the
crossed ladder diagram we have a crucial contribution coming from the other
Fock space sector corresponding to the pair production. This is in fact a
contribution that can be intuitively expected from the very nature of the
crossed diagram, but that has never been noticed before or even considered.
\ It is in this diagram that we can see manifestly the non-triviality of the
light-front vacuum without any external perturbating field acting on the
propagating bosons. We have already demonstrated that the non-trivial
light-front vacuum also does manifest in the propagation of two scalar
bosons when one of them is perturbed by an exernal background
electromagnetic field \cite{pairprod}.

In Classical Mechanics (CM), we characterize the state of a physical system
by a point in the four dimensional space-time, $(x_{i},t)$, where $x_{i}$
with $i=1,2,3$ and $t$ are respectively the space and the time coordinates.
If we parametrize the time coordinate, we can describe the same system in
the phase space $(q,p)$, with $q=(q_{1},q_{2},...,q_{n})$ indicating
collectively the degrees of freedom for such a system and $%
p=(p_{1},p_{2},...,p_{n})$ their respective conjugate momenta. Time
evolution of the system is then described by a line either in the space-time
or in the phase space $(q(t),p(t))$ which represents the set of values for
all degrees of freedom and respective conjugate momenta at a given instant
of time $t$. This time evolution is generated by Hamilton's equation. In
this strictly deterministic theory, the state of the system at a given time
is fixed by the initial conditions, as for example, at $t=0$.

In Quantum Mechanics (QM), or wave mechanics, on the other hand, the state
of a system is characterized by the wave function $\Psi (x,t)$, and its
evolution is governed by Schr\"{o}dinger's equation. The wave function must
be known at each point of the space-time $(x,t)$, so that it requires a
continuous infinite values to be described, with the wave function taking
the role of a field.

In a scattering process we have a wave function $\Psi (x,0)$, where we fix
the initial conditions at $t=0$, coming into a point which has a force field
or a particle, so that the typical question that arises is: Which is the
wave function representation $\Psi (x,t)$ after the interaction with the
scattering center?

In order to answer this question, we use the generalized Huygens principle 
\cite{19}. If a wave function $\Psi (x,t)$ is known at a given time $t$, it
is acceptable to assume that the wave function $\Psi (x^{\prime },t^{\prime
})$ that emerges from the scattering centre located at $(x,t)$ and
propagates from position $x$ to $x^{\prime }$ in a time $t^{\prime }-t$, be
proportional to the amplitude of the wave function $\Psi (x,t)$. The
proportionality contant is defined as $iG(x^{\prime },t^{\prime };x,t)$, so
that we have in a mathematical notation: 
\begin{equation*}
\Psi (x^{\prime },t^{\prime })=i\int d^{\,3}x\;G(x^{\prime },t^{\prime
};x,t)\,\Psi (x,t),
\end{equation*}
where $t^{\prime }>t$.

Here $\Psi (x^{\prime },t^{\prime })$ defined at position $x^{\prime}$ and
time $t^{\prime }$ is the wave function that emerges from the scattering
centre. The quantity $G(x^{\prime },t^{\prime };x,t)$ is called the Green's
function or propagator \cite{19}. Knowing this $G$ means to solve completely
the scattering problem. In other words, knowing the Green's function is
equivalent to solving of the Schr\"{o}dinger's equation.

Now, is it possible to describe a physical system in any space-time
hypersurface with initial conditions defined in a hypersurface different
from $t=0$? The answer to this question is positive and it was given long
ago in $1949$ by Dirac \cite{1}. In his work, he proposes three distinct
forms that could describe the dynamics of relativistic systems, two of which
do not use time to describe the dynamical properties of simple systems.
These different forms received different names: instant, point and front
forms. The instant form is the usual relativistic dynamics described in
terms of boundary conditions set at $t=0$ while the front form uses the
hyperplane of Minkowski space that contains the trajectory of light
(light-cone).

\section{Free boson}

The propagation of a free particle with spin zero in four dimensional
space-time is represented by the covariant Feynman propagator 
\begin{equation}
S(x^{\mu })=\int \frac{d^{4}k}{\left( 2\pi \right) ^{4}}\frac{ie^{-ik^{\mu
}x_{\mu }}}{k^{2}-m^{2}+i\varepsilon },  \label{2.1}
\end{equation}%
where the coordinate $x^{0}$ represents the time and $k^{0}$ the energy. We
are going to calculate this propagation in the light front, that is, for
times $x^{+}$.

A point in the four-dimensional space-time is defined by the set of numbers $%
(x^{0},x^{1},x^{2},x^{3})$, where $x^{0}$ is the time coordinate and $%
\mathbf{x}=(x^{1},x^{2},x^{3})$ is the three-dimensional vector with space
coordinates $x^{1}=x$, $x^{2}=y$ e $x^{3}=z$. Observe that we adopt here the
usual convention to take the speed of light as $c=1$.

In the light front, time and space coordinates are mixed up and we define
the new coordinates as follows: 
\begin{eqnarray}
x^{+} &=&x^{0}+x^{3}, \\
x^{-} &=&x^{0}-x^{3}, \\
\vec{x}^{\perp } &=&x^{1}\overrightarrow{i}+x^{2}\overrightarrow{j},
\end{eqnarray}
where $\overrightarrow{i}$ and $\overrightarrow{j}$ are unit vectors in the
direction of $x$ and $y$ coordinates respectively.

The null plane is defined by $x^{+}=0$, that is, this condition defines a
hyperplane that is tangent to the light-cone, the reason why many authors
call the hypersurface simply by light-cone.

The initial boundary conditions for the dynamics in the light front are
defined on this hyperplane. The axis $x^{+}$ is perpendicular to the plane $%
x^{+}=0$. Therefore a displacement of such hyperplane for $x^{+}>0$ is
analogous to the displacement of a plane in $t=0$ to $t>0$ of the
four-dimensional space-time. With this analogy, we recognize $x^{+}$ as the
time in the null plane.

We make the projection of the propagator for a boson in time associated to
the null plane rewriting the coordinates in terms of time coordinate $x^{+}$
and the position coordinates $(x^{-}$ and $\vec{x}_{\perp })$. With these,
the momenta are given by $k^{-}$, $k^{+}$ and $\vec{k}_{\perp }$, and
therefore we have 
\begin{equation}
S(x^{+})=\frac{1}{2}\int \frac{dk_{1}^{-}}{\left( 2\pi \right) }\frac{ie^{%
\frac{-i}{2}k_{1}^{-}x^{+}}}{k_{1}^{+}\left( k_{1}^{-}-\frac{k_{1\perp
}^{2}+m^{2}-i\varepsilon }{k_{1}^{+}}\right) }.  \label{2.2}
\end{equation}%
The Jacobian of the transformation $k^{0}$, $\vec{k}\rightarrow k^{-},k^{+},%
\vec{k}_{\perp }$ is equal to $\frac{1}{2}$ and $k^{+}$, $k_{\bot }$ are
momentum operators.

Evaluating the Fourier transform, we obtain 
\begin{equation}
\widetilde{S}(k^{-})=\int dx^{+}e^{\frac{i}{2}k^{-}x^{+}}S(x^{+}),
\label{2.3}
\end{equation}
where we have used 
\begin{equation}
\delta (\frac{k^{-}-k_{1}^{-}}{2})=\frac{1}{2\pi }\int dx^{+}e^{\frac{i}{2}%
\left( k^{-}-k_{1}^{-}\right) x^{+}},  \label{2.4}
\end{equation}
and the property of Dirac's delta ``function'' 
\begin{equation}
\delta \left( ax\right) =\frac{1}{a}\delta \left( x\right) ,  \label{2.5}
\end{equation}
and we get 
\begin{equation}
\widetilde{S}(k^{-})=\frac{i}{k^{+}\left( k^{-}-\frac{k_{\perp
}^{2}+m^{2}-i\varepsilon }{k^{+}}\right) },  \label{2.6}
\end{equation}
which describes the propagation of a particle forward to the future and of
an antiparticle backwards to the past. This can be oberved by the
denominator which hints us that for $x^{+}>0$ and $k^{+}>0$ we have the
particle propagating forward in time of the null plane. On the other hand,
for $x^{+}<0$ and $k^{+}<0$ we have an antiparticle propagating backwards in
time.

The Green funtion in the light front $G(x^{+})$ acting in the Fock space is
defined as the probability amplitude of the transition from the initial
state in the Fock space $\left| i\right\rangle $ to the final state $\left|
f\right\rangle $. Its Fourier transform is sometimes called resolvent for a
given Hamiltonian \cite{18}, however, here we call simply Fourier transform
for the Green function or even Green function itself.

In the case of a free boson, the Green function for the propagation of a
particle is defined by the operator 
\begin{equation}
G_{0}^{(1p)}(k^{-})=\frac{\theta (k^{+})}{k^{-}-k_{on}^{-}+i\varepsilon }\ ;
\label{2.7}
\end{equation}
where $k_{on}^{-}=\frac{k_{\perp }^{2}+m^{2}}{k^{+}}$ is the energy of the
particle. For the antiparticle propagation, we have: 
\begin{equation}
G_{0}^{(1a)}(k^{-})=\frac{\theta (-k^{+})}{k^{-}-k_{on}^{-}-i\varepsilon }\ .
\label{2.8}
\end{equation}

We can see that the difference between the Green functions in (\ref{2.7})
and (\ref{2.8}) for the propagator in the light front is the absence of the
imaginary (complex) number $i$ and of the factor of phase space $k^{+}$
which appears in (\ref{2.6}).

The operator defined by (\ref{2.7}) is the Green function of 
\begin{equation}
\left( k^{-}-k_{on}^{-}\right) \left(
G_{0}^{(1p)}(k^{-})+G_{0}^{(1a)}(k^{-})\right) =1\ .  \label{2.9}
\end{equation}

The Feynman propagator is then rewritten as: 
\begin{equation}
S(k^{\mu })=\frac{i}{k^{+}}G_{0}^{(1p)}(k^{-})-\frac{i}{|k^{+}|}%
G_{0}^{(1a)}(k^{-})=\frac{i}{k^{+}(k^{-}-\frac{k_{\perp
}^{2}+m^{2}-i\varepsilon }{k^{+}})}\ .  \label{2.10}
\end{equation}

\section{Green function of two bosons}

Our aim in this chapter is to study the two body Green function in the
``ladder'' approximation for the dynamics defined in the light front. Within
this treatment, we are not going to deal with perturbative corrections that
can be decomposed into one body problem.

Our interest is to define in the light front, the interaction between two
bodies mediated by the interchange of a particle and obtain the correction
to the two body Green function originated in this interaction.

For this purpose, we use a bosonic model for which the interaction
Lagrangian is defined as: 
\begin{equation}
\mathcal{L}_{I}=g\phi _{1}^{\ast }\phi _{1}\sigma +g\phi _{2}^{\ast }\phi
_{2}\sigma ,  \label{lagr}
\end{equation}%
where the bosons $\phi _{1}$ and $\phi _{2}$ have equal mass $m$ and the
intermediate boson, $\sigma $, has the mass $m_{\sigma }$. The coupling
constant is $g$.

Taking from Dirac's idea \cite{1} of representing the dynamics of a quantum
system in the light front in time $x^{+}=t+z$, we derive in this chapter the
two body Green function or covariant propagator which describes the
evolution of the system from a hypersurface $x^{+}=\mathrm{constant}$ to
another one. The Green function in the light front is the probability
amplitude for a initial state in $x^{+}=0$ evolving to a final state in $%
x^{+}>0$, where the evolution operator is defined by the Hamiltonian in the
light front \cite{2}. Sometimes we call the resolvent $\left( Z-H\right)
^{-1}$ as the Green function, too \cite{36}.

The two body Green function in the light front includes the propagation of
intermediate states with any number of particles.

We start our discussion evaluating the second order correction to the
coupling constant associated with the propagator. We define the matrix
element for the interaction and so we obtain the correction to the Green
function in the light front. Then we evaluate the correction to the Green
function to the fourth order in the coupling constant, where we use the
technique of factorizing the energy denominators, which is important to
identify the global propagation of four bodies after the integration in the
energies $k^{-}$. We discuss the generalization of this technique.

We show how to get the non perturbative Green function from the set of
hierarchical equations for the Green function in the \textquotedblleft
ladder\textquotedblright\ approximation. This corresponds to the truncation
of the Fock space in the light front, such that, only states with two bosons 
$\phi _{1}$ and $\phi _{2}$ are allowed, with no restriction as to the
number of intermediate bosons $\sigma $. We discuss how to build a
systematic approximation to the kernel of the integral equation for the two
body Green function as a function of the number of particles in the
intermediate Fock state and the power in the coupling constant. A consistent
truncation can be carried out and in the lowest order, this brings to the
Weinberg's equation for the bound state \cite{10}.

The propagator for the diagrams in interaction can be obtained of the
functional generator (\ref{green2}) and the are shown in the next sections.

\section{One boson exchange}

The perturbative correction to the propagator of two bodies in $O(g^{2})$
comes from the interchange of a virtual intermediate boson, given by: 
\begin{eqnarray}
\Delta S_{g_{S}^{2}}^{(2)}(x^{+}) &=&(ig)^{2}\int d\overline{x}_{1}^{+}d%
\overline{x}_{2}^{+}S_{3}(x^{+}-\overline{x}_{1}^{+})S_{4}(x^{+}-\overline{x}%
_{2}^{+})  \notag \\
&&\times S_{\sigma }(\overline{x}_{1}^{+}-\overline{x}_{2}^{+})S_{1}(%
\overline{x}_{1}^{+})S_{2}(\overline{x}_{2}^{+})\ .  \label{3.2}
\end{eqnarray}

The propagator of a particle $S_{i}(x^{+})$ is defined in (\ref{2.2}). The
intermediate boson $\sigma $ propagates between $\overline{x}_{1}^{+}$ and $%
\overline{x}_{2}^{+}$. The indices $1$ and $2$ in the particle propagators
label initial states whereas $3$ and $4$ do so for the final states.

\section{Two boson exchange (ladder diagram):}

In this section, we evaluate the perturbative correction to the propagator
of two bosons up to the fourth order in the coupling constant, using the
same method of the previous section.This corresction to the propagator
constains terms obtained from the iteration of the Bethe-Salpeter equation
in the ``ladder'' approximation up to the second order.

As we have a correction for the two boson propagator to the fourth order in
the coupling constant, we have eigth propagators in the \textquotedblleft
ladder\textquotedblright\ diagram, two of which are propagators of the boson
associated with the interation, identified as $k_{\sigma }$ e $k_{\sigma
^{\prime }}$. We also make use of a lower index to define the particle
momenta that propagate.

The perturbative correction to the Feynman propagator for two bodies up to
order $g^{4}$ in the ``ladder'' approximation is given by: 
\begin{eqnarray}
\Delta S_{g^{4}}^{(2)}(x^{+}) &=&(ig)^{4}\int d\overline{x}_{1}^{+}d%
\overline{x}_{2}^{+}d\overline{x}_{3}^{+}d\overline{x}_{4}^{+}S_{5}(x^{+}-%
\overline{x}_{3}^{+})S_{6}(x^{+}-\overline{x}_{4}^{+})S_{\sigma ^{\prime }}(%
\overline{x}_{3}^{+}-\overline{x}_{4}^{+})\times  \notag \\
&&S_{3}(\overline{x}_{3}^{+}-\overline{x}_{1}^{+})S_{4}(\overline{x}_{4}^{+}-%
\overline{x}_{2}^{+})\ S_{\sigma }(\overline{x}_{1}^{+}-\overline{x}%
_{2}^{+})S_{1}(\overline{x}_{1}^{+})S_{2}(\overline{x}_{2}^{+})\ .
\label{3.11}
\end{eqnarray}

Each propagator of the previous expression, is defined by (\ref{2.2}).
Therefore we substitute them in (\ref{3.11}) and integrate in $\overline{x}%
_{1},\overline{x}_{2},\overline{x}_{3}$ and $\overline{x}_{4}$. With this,
there emerges four Dirac deltas in $k_{j}^{-}(j=1,...,\sigma ,\sigma
^{\prime })$, which correspond to the conservation of $k^{-}$ in each
vertex. So with these, eliminating four integrations in $k^{-}$, we are left
with an exponencial $\mathrm{e}^{\frac{-i}{2}\left(
k_{5}^{-}+k_{6}^{-}\right) x^{+}}$. Evaluating the Fourier transform and
integrating in $x^{+}$, this exponencial makes up another Dirac delta which
later on, when performing integration in $k_{3}^{-}$ results in the law of
energy conservation for the system, given by, 
\begin{equation*}
K^{-}=k_{5}^{-}+k_{6}^{-}=k_{1}^{-}+k_{2}^{-},
\end{equation*}%
where $k_{5}^{-}$ and $k_{6}^{-}$ are light front energies for the final
state particles and $k_{1}^{-}$ and $k_{2}^{-}$ for the initial ones.

We note that in this ladder diagram there is no pair production and the two
Fock spaces remain separate; the vacuum is trivial.

\section{Cross Ladder Diagram}

This diagram brings about a new feature which was not present in the
previous diagram just considered. Here we come across not only with all
those diagrams that involve the propagation of information to future times,
but note one particular diagram that bears pair production, i.e., there is
one diagram which has intrinsic propagation of information to the past, thus
mingling the two sectorized Fock spaces of solutions. This diagram, as far
as we known, has not being considered in the literature before.

The correction to the two boson propagator coming from the process of two \
"crossed " $\sigma $ bosons is represented by the Feynman diagram depicted
in figure (\ref{f4.1a}).

\FRAME{ftbpFU}{3.0743in}{1.1748in}{0pt}{\Qcb{Exchange of two \ "crossed " $%
\protect\sigma $ bosons}}{\Qlb{f4.1a}}{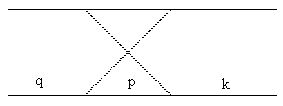}{\special{language
"Scientific Word";type "GRAPHIC";maintain-aspect-ratio TRUE;display
"USEDEF";valid_file "F";width 3.0743in;height 1.1748in;depth
0pt;original-width 2.264in;original-height 0.8485in;cropleft "0";croptop
"1";cropright "1";cropbottom "0";filename 'fg6r4.jpg';file-properties
"XNPEU";}}

Explicitly this correction to the propagator is written down in terms of the
one boson propagators and is given by the following equation%
\begin{eqnarray}
\Delta S_{\times }(x^{+}) &=&(ig)^{4}\int d\overline{x}_{1}^{+}d\overline{x}%
_{2}^{+}d\overline{x}_{3}^{+}d\overline{x}_{4}^{+}S_{3}(x^{+}-\overline{x}%
_{2}^{+})\times  \notag \\
&&S_{\sigma }(\overline{x}_{2}^{+}-\overline{x}_{3}^{+})S_{2}(\overline{x}%
_{2}^{+}-\overline{x}_{1}^{+})S_{6}(x^{+}-\overline{x}_{4}^{+})\times  \notag
\\
&&S_{\sigma }(\overline{x}_{4}^{+}-\overline{x}_{1}^{+})S_{5}(\overline{x}%
_{4}^{+}-\overline{x}_{3}^{+})\times  \notag \\
&&S_{1}(\overline{x}_{1}^{+})S_{4}(\overline{x}_{3}^{+}),  \label{1c}
\end{eqnarray}%
and after Fourier transform we have

\begin{eqnarray}
\Delta \widetilde{S}_{\times }(K^{-}) &=&\frac{\left( ig\right) ^{4}i^{8}}{%
2^{3}\left( 2\pi \right) ^{3}}\int dk^{-}dp^{-}dq^{-}\frac{%
dp^{+}d^{2}p_{\perp }}{k^{+}p^{+}(k-p)^{+}(p-q)^{+}(K-k-q+p)^{+}}\times 
\notag \\
&&\frac{1}{\left( K-q\right) ^{+}\left( q-p\right) ^{+}\left( k-p\right) ^{+}%
}\times  \notag \\
&&\frac{1}{\left( k^{-}-\frac{k_{\perp }^{2}+m^{2}-i\varepsilon }{k^{+}}%
\right) }\frac{1}{\left( p^{-}-\frac{k_{\perp }^{2}+m^{2}-i\varepsilon }{%
p^{+}}\right) }\times  \notag \\
&&\frac{1}{\left( q^{-}-\frac{k_{\perp }^{2}+m^{2}-i\varepsilon }{q^{+}}%
\right) }\frac{1}{\left( K^{-}-k^{-}-\frac{\left( K-k\right) _{\perp
}^{2}+m^{2}-i\varepsilon }{\left( K-k\right) ^{+}}\right) }  \notag \\
&&\frac{1}{\left( K^{-}-q^{-}-\frac{\left( K-q\right) _{\perp
}^{2}+m^{2}-i\varepsilon }{\left( K-q\right) ^{+}}\right) }\frac{1}{\left(
k^{-}-p^{-}-\frac{\left( k-p\right) _{\perp }^{2}+m_{\sigma
}^{2}-i\varepsilon }{\left( k-p\right) ^{+}}\right) }  \notag \\
&&\frac{1}{\left( K^{-}-k^{-}-q^{-}+p^{-}-\frac{\left( K-k-q+p\right)
_{\perp }^{2}+m^{2}-i\varepsilon }{\left( K-k-q+p\right) ^{+}}\right) } 
\notag \\
&&\frac{1}{\left( q^{-}-p^{-}-\frac{\left( q-p\right) _{\perp
}^{2}+m_{\sigma }^{2}-i\varepsilon }{\left( q-p\right) ^{+}}\right) },
\label{2c}
\end{eqnarray}

For $K^{+}>0$, the regions of integration in $p^{+}$ which define the
position of poles in the complex $p^{-}$ are:

a) $0<q^{+}<p^{+}<k^{+}<K^{+}$

b) $0<k^{+}<p^{+}<q^{+}<K^{+}$

c) $0<p^{+}<q^{+}<k^{+}<K^{+}$

d) $0<p^{+}<k^{+}<q^{+}<K^{+}$

e) $0<k^{+}<q^{+}<p^{+}<K^{+}$

f) $0<q^{+}<k^{+}<p^{+}<K^{+}$

For regions \textquotedblleft $c$ \textquotedblright\ and \textquotedblleft $%
d$ \textquotedblright\ we use the method of partial fracioning twice to
integrate in $p^{-}$ \cite{14}; for \textquotedblleft $a$ \textquotedblright
, \textquotedblleft $b$ \textquotedblright , \textquotedblleft $e$
\textquotedblright\ and \textquotedblleft $f$ \textquotedblright\ this is
not necessary; for regions \textquotedblleft $e$ \textquotedblright\ and
\textquotedblleft $f$ \textquotedblright\ the integration in $p^{-}$
vanishes. \ The diagrams for each of these regions are shown in figure (\ref%
{f4.2a}).

\FRAME{ftbpFU}{3.9087in}{3.203in}{0pt}{\Qcb{Crossed ladder diagrams time
ordered in $x^{+}>0$.}}{\Qlb{f4.2a}}{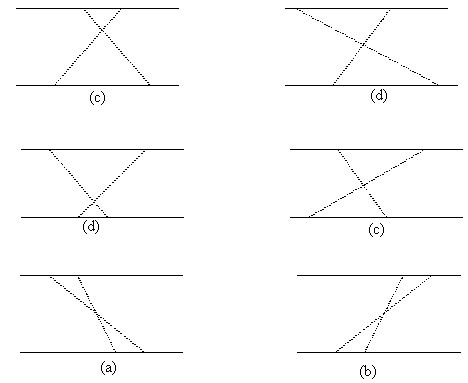}{\special{language
"Scientific Word";type "GRAPHIC";maintain-aspect-ratio TRUE;display
"USEDEF";valid_file "F";width 3.9087in;height 3.203in;depth
0pt;original-width 3.7443in;original-height 3.0643in;cropleft "0";croptop
"1";cropright "1";cropbottom "0";filename 'fg7r4.jpg';file-properties
"XNPEU";}}

After performing the analytic integrations in $k^{-},p^{-}$ and $q^{-}$ we
have 
\begin{equation}
\Delta \widetilde{S}_{\times }(K^{-})=\Delta \widetilde{S}_{\times
}^{a}(K^{-})+\Delta \widetilde{S}_{\times }^{b}(K^{-})+\Delta \widetilde{S}%
_{\times }^{c}(K^{-})+\Delta \widetilde{S}_{\times }^{d}(K^{-})  \label{3c}
\end{equation}%
where 
\begin{eqnarray}
\Delta \widetilde{S}_{\times }^{a}(K^{-}) &=&(ig)^{4}\int \frac{%
idp^{+}d^{2}p_{\perp }\theta (k^{+}-p^{+})\theta (p^{+}-q^{+})}{%
2k^{+}(K-k)^{+}\left( K^{-}-\frac{k_{\perp }^{2}+m^{2}-i\varepsilon }{k^{+}}-%
\frac{(K-k)_{\perp }^{2}+m^{2}-i\varepsilon }{(K-k)^{+}}\right) }  \notag \\
&&\frac{1}{2p^{+}(k-p)^{+}(p-q)^{+}(K-k-q+p)^{+}}\times   \notag \\
&&\frac{i}{\left( K^{-}-\frac{p_{\perp }^{2}+m^{2}-i\varepsilon }{p^{+}}-%
\frac{(K-k)_{\perp }^{2}+m^{2}-i\varepsilon }{(K-k)^{+}}-\frac{(k-p)_{\perp
}^{2}+m^{2}-i\varepsilon }{(k-p)^{+}}\right) }\times   \notag \\
&&\frac{i}{\left( K^{-}-\frac{(p-q)_{\perp }^{2}+m_{\sigma
}^{2}-i\varepsilon }{(p-q)^{+}}-\frac{(k-p)_{\perp }^{2}+m_{\sigma
}^{2}-i\varepsilon }{(k-p)^{+}}-\frac{(K-k)_{\perp }^{2}+m^{2}-i\varepsilon 
}{(K-k)^{+}}-\frac{q_{\perp }^{2}+m^{2}-i\varepsilon }{q^{+}}\right) }\times 
\notag \\
&&\frac{i}{\left( K^{-}-\frac{q_{\perp }^{2}+m^{2}-i\varepsilon }{q^{+}}-%
\frac{(k-p)_{\perp }^{2}+m_{\sigma }^{2}-i\varepsilon }{(k-p)^{+}}-\frac{%
(K-k-q+p)_{\perp }^{2}+m^{2}-i\varepsilon }{(K-k-q+p)^{+}}\right) }  \notag
\\
&&\frac{i}{2q^{+}(K-q)^{+}\left( K^{-}-\frac{q_{\perp
}^{2}+m^{2}-i\varepsilon }{q^{+}}-\frac{(K-q)_{\perp
}^{2}+m^{2}-i\varepsilon }{(K-q)^{+}}\right) },  \label{4c}
\end{eqnarray}%
and 
\begin{equation}
\Delta \widetilde{S}_{\times }^{b}(K^{-})=\Delta \widetilde{S}_{\times
}^{a}(K^{-})[k\leftrightarrow q],  \label{5c}
\end{equation}%
shown in figures (\ref{f4.2a}a) and (\ref{f4.2a}b) respectively.

Regions \textquotedblright $c$\textquotedblright\ and \textquotedblright $d$%
\textquotedblright\ contribute to the propagator correction as 
\begin{eqnarray}
\Delta \widetilde{S}_{\times }^{c}(K^{-}) &=&(ig)^{4}\int \frac{%
idp^{+}d^{2}p_{\perp }\theta (q^{+}-p^{+})\theta (k^{+}-p^{+})}{%
2k^{+}(K-k)^{+}\left( K^{-}-\frac{k_{\perp }^{2}+m^{2}-i\varepsilon }{k^{+}}-%
\frac{(K-k)_{\perp }^{2}+m^{2}-i\varepsilon }{(K-k)^{+}}\right) }  \notag \\
&&\frac{1}{2p^{+}(k-p)^{+}(p-q)^{+}(K-k-q+p)^{+}}\times  \notag \\
&&\frac{i}{\left( K^{-}-\frac{p_{\perp }^{2}+m^{2}-i\varepsilon }{p^{+}}-%
\frac{(q-p)_{\perp }^{2}+m_{\sigma }^{2}-i\varepsilon }{(q-p)^{+}}-\frac{%
(K-k-q+p)_{\perp }^{2}+m^{2}-i\varepsilon }{(K-k-q+p)^{+}}-\frac{%
(k-p)_{\perp }^{2}+m_{\sigma }^{2}-i\varepsilon }{(k-p)^{+}}\right) }\times 
\notag \\
&&\widetilde{S^{\prime }}_{\times }^{c}\frac{i}{2q^{+}(K-q)^{+}\left( K^{-}-%
\frac{q_{\perp }^{2}+m^{2}-i\varepsilon }{q^{+}}-\frac{(K-q)_{\perp
}^{2}+m^{2}-i\varepsilon }{(K-q)^{+}}\right) },  \label{6c}
\end{eqnarray}%
where

\begin{eqnarray}
\widetilde{S^{\prime }}_{\times }^{c} &=&\frac{i}{\left( K^{-}-\frac{%
q_{\perp }^{2}+m^{2}-i\varepsilon }{q^{+}}-\frac{(k-p)_{\perp
}^{2}+m_{\sigma }^{2}-i\varepsilon }{(k-p)^{+}}-\frac{(K-k-q+p)_{\perp
}^{2}+m^{2}-i\varepsilon }{(K-k-q+p)^{+}}\right) }\times   \notag \\
&&\frac{i}{\left( K^{-}-\frac{k_{\perp }^{2}+m^{2}-i\varepsilon }{k^{+}}-%
\frac{(K-k-q+p)_{\perp }^{2}+m_{\sigma }^{2}-i\varepsilon }{(K-k-q+p)^{+}}-%
\frac{(q-p)_{\perp }^{2}+m^{2}-i\varepsilon }{(q-p)^{+}}\right) }\times  
\notag \\
&&\frac{i}{\left( K^{-}-\frac{k_{\perp }^{2}+m^{2}-i\varepsilon }{k^{+}}-%
\frac{(q-p)_{\perp }^{2}+m_{\sigma }^{2}-i\varepsilon }{(q-p)^{+}}-\frac{%
(K-k-q+p)_{\perp }^{2}+m^{2}-i\varepsilon }{(K-k-q+p)^{+}}\right) }\times  
\notag \\
&&\frac{i}{\left( K^{-}-\frac{p_{\perp }^{2}+m^{2}-i\varepsilon }{p^{+}}-%
\frac{(q-p)_{\perp }^{2}+m_{\sigma }^{2}-i\varepsilon }{(q-p)^{+}}-\frac{%
(K-q)_{\perp }^{2}+m^{2}-i\varepsilon }{(K-q)^{+}}\right) }.  \label{7c}
\end{eqnarray}%
The perturbative correction to the two boson propagator in Eq.(\ref{6c}) is
represented by diagrams indicated in figure (\ref{f4.2a}c). The correction
represented by diagrams in figure (\ref{f4.2a}d) is given by: 
\begin{equation}
\Delta \widetilde{S}_{\times }^{d}(K^{-})=\Delta \widetilde{S}_{\times
}^{c}(K^{-})\left[ k\leftrightarrow K-k,p\leftrightarrow
K-k-q+p,q\leftrightarrow K-q\right] .  \label{8c}
\end{equation}

\subsection{Antiparticle contribution}

Next we deduce the antiparticle contribution to the crossed ladder diagram.
This contribution happens for $p^{+}<0$ and $K^{+}-k^{+}-q^{+}+p^{+}<0$. Let
us analyse the first case, $p^{+}<0.$

The region for the $"+"$ component momentum that allows the pole positioned
in both hemispheres of the complex $p^{-}$ plane, and therefore giving
non-vanishing residue, are $-k^{+}<p^{+}<0$ and $\left\vert p^{+}\right\vert
+k^{+}+q^{+}<K^{+}$. So, the result for the momentum integration in $"-"$
component for $0<q^{+}<K^{+}$ and $0<k^{+}<K^{+}$ which correspond to the
non-vanishing results for integrations in $k^{-}$ and $q^{-}$ for $%
-k^{+}<p^{+}<0$, is given by the diagram depicted in figure (\ref{f12a}) and
the result is given in (\ref{9c}):

\FRAME{ftbpFU}{2.6999in}{1.0021in}{0pt}{\Qcb{Pair creation process
contributing to the crossed ladder diagram.}}{\Qlb{f12a}}{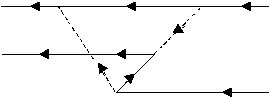}{\special%
{language "Scientific Word";type "GRAPHIC";maintain-aspect-ratio
TRUE;display "USEDEF";valid_file "F";width 2.6999in;height 1.0021in;depth
0pt;original-width 2.1602in;original-height 0.7846in;cropleft "0";croptop
"1";cropright "1";cropbottom "0";filename 'f12a.jpg';file-properties
"XNPEU";}} 
\begin{eqnarray}
\Delta \widetilde{S}_{\times }^{ant}(K^{-}) &=&\frac{(ig)^{4}}{8}\int
dp^{+}d^{2}p_{\perp }\frac{\theta (k^{+}+\left\vert p^{+}\right\vert )\theta
(q^{+}+\left\vert p^{+}\right\vert )}{k^{+}\left\vert p^{+}\right\vert
q^{+}(\left\vert p\right\vert +k)^{+}(q+\left\vert p\right\vert )^{+}}\times 
\notag \\
&&\frac{\theta (K^{+}-\left\vert p^{+}\right\vert -q^{+}-k^{+})}{\left(
K-k\right) ^{+}\left( K-q\right) ^{+}(K-\left\vert p\right\vert -q-k)^{+}}%
\times   \notag \\
&&\frac{i}{(K^{-}-K_{0}^{-})}\frac{i}{(K^{-}-Q_{0}^{-})}\frac{i}{%
(K^{-}-T^{-})}\times   \notag \\
&&\frac{i}{(K^{-}-J_{a}^{-})}\frac{i}{(K^{-}-T^{\prime -})}+  \label{9c} \\
+[k &\leftrightarrow &q]+[k\rightarrow K-k,q\rightarrow K-q]+[k\rightarrow
K-q,q\rightarrow K-k],  \notag
\end{eqnarray}%
where 
\begin{eqnarray}
T^{-} &=&\frac{(p-q)_{\bot }^{2}+m_{\sigma }^{2}}{q^{+}+\left\vert
p^{+}\right\vert }+\frac{(K+p-q-k)_{\bot }^{2}+m^{2}}{K^{+}-k^{+}-q^{+}-%
\left\vert p^{+}\right\vert }+  \notag \\
&&+\frac{k_{\bot }^{2}+m^{2}}{k^{+}},  \notag \\
J_{a}^{-} &=&\frac{q_{\bot }^{2}+m^{2}}{q^{+}}+\frac{(K-k-q+p)_{\bot
}^{2}+m^{2}}{K^{+}-k^{+}-q^{+}-\left\vert p^{+}\right\vert }+  \notag \\
&&+\frac{p_{\bot }^{2}+m^{2}}{\left\vert p^{+}\right\vert }+\frac{k_{\bot
}^{2}+m^{2}}{k^{+}},  \notag \\
T^{\prime -} &=&\frac{(q-p)_{\bot }^{2}+m_{\sigma }^{2}}{q^{+}+\left\vert
p^{+}\right\vert }+\frac{(K-k-w+p)_{\bot }^{2}+m^{2}}{K^{+}-k^{+}-q^{+}-%
\left\vert p^{+}\right\vert }+  \notag \\
&&+\frac{q_{\bot }^{2}+m^{2}}{q^{+}}.  \label{10c}
\end{eqnarray}%
\qquad 

The four body propagator, $J_{a}^{-},$\ (subindex $a$ for antiparticle) has
a propagation to past in the null-plane of an antiparticle with $p^{+}<0$.
At instant $\overline{x}_{2}^{+}>0$ the pair particle-antiparticle is
produced by the $\sigma $ intermediate boson, then the antiparticle
encounters a particle of momentum $k^{+}>0$, and is anihilated and the
production of a $\sigma $ boson with momentum $\left\vert p^{+}\right\vert
+k^{+}>0$ which continues to propagate into the future of the null-plane.

\section{Conclusions}

In our previous work we have demonstrated the pair production in the
propagation of two scalar bosons in which one of them interacts with an
external electromagnetic field. There, the propagation of a single boson
interacting with the external field does not produce pair since this event
is isolated; the second propagating boson which is the \ "spectator " boson
(the one that sees the event happening) is crucial to the process to occur.

In the case of the ladder diagrams, the usual one does not produce pair too;
only the crossed ladder diagram does. This is due to the fact that
propagating intermediate bosons in crossed ladder diagrams has to have
null-plane time ordered operators linked with causality playing its crucial
role.

One very important observation that we pinpoint here is that in the
light-front dynamics, the light-front coordinates $x^{+}$ and $x^{-}$ are
not simple $45^{0}$ rotation of the usual four-dimensional space-time
coordinates $x^{0}$ and $x^{3}$. It is rather a rotation constrained by the
condition of light-likeness of the vector, i.e., $x^{2}=0$, which renders
the components $x^{+}$ and $x^{-}$ as being \textsc{not} linearly
independent components; they do not serve as base vectors to generate the
four-dimensional space-time. Moreover in the momentum integration, $k^{+}$
and $k^{-}$ we have the same linearly dependence, here the difference being
that the dispersion relation is the constraint, either $k^{2}=0$ for
massless particles or $k^{2}=m^{2}$ for massive particles (relation known as
the Einstein relation).

We believe that most of the (if not all of them...) pathologies emerging in
the light-front treatment of the physical processes up to now have been
plagued by this simple misunderstanding and lack of perception. Our
calculation has also been victimized by this lack of perception before. In
reviewing the pathological behaviour of the process, for which no physical
argumentation has been proposed to cure it, we were led to analyse more
carefully the constraint defined by the dispersion relation. From it emerged
the paradigm that in the light-front energy $k^{-}$ and longitudinal
momentum component $k^{+}$ are bound together in their signs; whether
positive or negative, both of them bear the same sign as the other one.
However, when we split the two and write energy in terms of momentum
components, the longitudinal \ "$+$ "-component of the momentum is carrying
relevant physical information which can be lost by treating them as
independent variables - which they certainly are not.

It is because of the above referred paradigm that up to now, most of the
calculations done in the light front have always neglected or omitted the
relevant contributions of the Fock space defined by the other sector,
namely, the region of negative $k^{+}.$

\appendix{}

\section{Generating Functional}

Using the concept of generating functional $Z\left[ J\right] $, or
vacuum-vacuum transition amplitude in the presence of an external source $%
J\left( x\right) $, we write: 
\begin{eqnarray}
Z\left[ J\right] &=&\int \mathcal{D}\phi \exp \left\{ i\int d^{4}x\left[ 
\pounds \left( \phi \right) +J\left( x\right) +\frac{i\varepsilon }{2}\phi %
\right] \right\}  \label{3.1.1} \\
&\varpropto &\left\langle 0,\infty \right. \left\vert 0,-\infty
\right\rangle ^{J}  \notag
\end{eqnarray}%
where $\pounds (\phi )$ is the free scalar field Lagrangean (or Klein-Gordon
field Lagrangean).

\begin{equation}
\pounds \left( \phi \right) =\frac{1}{2}\left( \partial _{\mu }\phi \partial
^{\mu }\phi -m^{2}\phi ^{2}\right)  \label{3.1.2}
\end{equation}%
Using now the identity: 
\begin{equation}
\partial _{\mu }\left( \phi \partial ^{\mu }\phi \right) =\partial _{\mu
}\phi \partial ^{\mu }\phi +\phi \partial _{\mu }\partial ^{\mu }\phi
\label{3.1.3}
\end{equation}%
substituting this expression above into the Lagrangean (\ref{3.1.2}) which
must be integrated over $d^{4}x$ to calculate the generating functional. In
this integration, the total derivative term (or divergent) does not
contribute, due to Gauss ' theorem in four dimensions and considering that
the field $\phi $ vanishes at infinity we have from (\ref{3.1.3}). 
\begin{equation}
\int \partial _{\mu }\phi \partial ^{\mu }\phi d^{4}x=-\int \phi \square
\phi d^{4}x  \label{3.1.4}
\end{equation}

We can then write the generating functional for the Klein-Gordon field
without interaction, which we denote as $Z_{0}\left[ J\right] $: 
\begin{equation}
Z_{0}\left[ J\right] =\int \mathcal{D}\phi \exp \left\{ -i\int \left[ \frac{1%
}{2}\phi \left( \square +m^{2}-i\varepsilon \right) \phi -J\phi \right]
d^{4}x\right\}  \label{3.1.5}
\end{equation}%
which after a simple calculation can be rewritten as: 
\begin{equation}
Z_{0}\left[ J\right] =N\exp \left[ -\frac{1}{2}\int J\left( x\right) \Delta
_{F}\left( x-y\right) J\left( y\right) dxdy\right]  \label{3.1.6}
\end{equation}%
where $dx$ and $dy$ stand for $d^{4}x$ and $d^{4}y$ respectively and the
normalization factor is 
\begin{equation*}
N=\int \mathcal{D}\phi \exp \left[ -\frac{1}{2}\int \phi \left( \square
+m^{2}-i\varepsilon \phi dx\right) \right]
\end{equation*}

We note that $N$ does not depend on the external source $J\left( x\right) $
and since we are interested only in normalized transition amplitudes, we can
write:%
\begin{equation*}
Z_{0}\left[ J\right] =\frac{\left\{ \mathcal{D}\phi \exp \left\{ -i\int %
\left[ \frac{1}{2}\phi \left( \square +m^{2}-i\varepsilon \right) \phi
-J\phi \right] dx\right\} \right\} }{\left\{ \int \mathcal{D}\phi \exp
\left\{ -i\int \frac{1}{2}\phi \left( \square +m^{2}-i\varepsilon \right)
\phi dx\right\} \right\} }
\end{equation*}%
\begin{equation}
Z_{0}\left[ J\right] =\exp \left[ -\frac{i}{2}\int J\left( x\right) \Delta
_{F}\left( x-y\right) J\left( y\right) dxdy\right]  \label{3.1.7}
\end{equation}%
where $\Delta _{F}\left( x\right) $ is the Feynmam propagator for the free
scalar field and which has the following Fourier transformed representation: 
\begin{equation}
\Delta _{F}\left( x\right) =\frac{1}{\left( 2\pi \right) ^{4}}\int d^{4}k%
\frac{e^{-ikx}}{k^{2}-m^{2}+i\varepsilon }  \label{3.1.8}
\end{equation}%
and is the solution for the equation 
\begin{equation}
\left( \square +m^{2}-i\varepsilon \right) \Delta _{F}\left( x\right)
=-\delta ^{4}\left( x\right)  \label{3.1.9}
\end{equation}

The Green 's functions are the expectation values of the time ordered field
operators in the vacuum and can be written in terms of functional
derivatives of the generating functional $Z_{0}\left[ J\right] $, that is, $%
G\left( x_{1},...,x_{n}\right) =\left\langle 0\right\vert T\left( \phi
\left( x_{1}\right) ...\phi \left( x_{n}\right) \right) \left\vert
0\right\rangle $ which is the $n$-point Green 's function of the theory,
where 
\begin{equation}
\left\langle 0\right\vert T\left( \phi \left( x_{1}\right) ...\phi \left(
x_{n}\right) \right) \left\vert 0\right\rangle =\frac{1}{i^{n}}\left. \frac{%
\delta ^{n}Z_{0}\left[ J\right] }{\delta J\left( x_{1}\right) ...\delta
J\left( x_{n}\right) }\right\vert _{J=0}.  \label{green2}
\end{equation}

Green's functions in field theory are extremely important because they are
intimately related to the elements of the scattering matrix $S$ through
which we can calculate quantities measured directly from the experiments
such as scattering processes where cross section for a particular reaction
is measured, or decay of a particle into two or more where we can measure
the half-lives of the particles involved, etc.

The propagator is associated to the Green's function equation as: 
\begin{equation}
G(t-t^{\prime })=-iS(t-t^{\prime }).  \label{gpro}
\end{equation}

The Green 's function or the propagator describes completely the time
evoltuion for the quantum system. In this present case we are using the
propagator for \textquotedblleft future times\textquotedblright . We could
also have defined the propagator \textquotedblleft
backwards\textquotedblright\ in time.

\textit{Acknowledgments:} J.H.O.Sales thanks the hospitality of Instituto de
F\'{\i}sica Te\'{o}rica/UNESP, where part of this work has been
done.A.T.Suzuki thanks the kind hospitality of Physics Department, North
Carolina Staste University where part of this work has been done and
gratefully acknowledges partial support from CNPq (Bras\'{\i}lia) in the
earlier stages of this work, then superseded by a grant from CAPES (Bras%
\'{\i}lia).

\end{document}